\documentclass[conference]{IEEEtran}

\usepackage{epsfig,makeidx,color}
\usepackage{graphicx}
\usepackage{amsbsy}
\usepackage{amsmath}
\usepackage{amssymb}
\usepackage{euscript,verbatim}

\newtheorem{thm}{Theorem}

\newcommand{\F}{\mathbf{F}}

\begin{document}

\title{Node Repair for Distributed Storage Systems over Fading Channels }

\author{
  \IEEEauthorblockN{David Karpuk, Camilla Hollanti, Amaro Barreal}
  \IEEEauthorblockA{Dept. of Mathematics and Systems Analysis\\
    Aalto University\\
    P.O.\ Box 11100\\
    FI-00076 Aalto, Finland\\
    email: \{david.karpuk, camilla.hollanti, amaro.barreal\}@aalto.fi} 
}

\maketitle

\begin{abstract}

Distributed storage systems and associated storage codes can efficiently store a large amount of data while ensuring that data is retrievable in case of node failure.  The study of such systems, particularly the design of storage codes over finite fields, assumes that the physical channel through which the nodes communicate is error-free.  This is not always the case, for example, in a wireless storage system.

We study the probability that a subpacket is repaired incorrectly during node repair in a distributed storage system, in which the nodes communicate over an AWGN or Rayleigh fading channels.  The asymptotic probability (as SNR increases) that a node is repaired incorrectly is shown to be completely determined by the repair locality of the DSS and the symbol error rate of the wireless channel.   Lastly, we propose some design criteria for physical layer coding in this scenario, and use it to compute optimally rotated QAM constellations for use in wireless distributed storage systems.

\end{abstract}


\IEEEpeerreviewmaketitle

\section{Introduction}\label{introduction}

In a distributed storage system (DSS), data is distributed among several storage nodes in a network, in such a way that the data is made robust against node failures by introducing redundancy.  There are many invariants of a DSS which measure its effectiveness.  Among the most important are repair bandwidth, storage, and \emph{repair locality}, which is the number of nodes a newcomer node needs to contact to repair the system.  The importance of repair locality was discovered independently in \cite{papadimlots}, \cite{oggierdatta}, \cite{microsoftlocality} and linear codes over finite fields, known as \emph{locally repairable codes} in \cite{papadim} and \emph{self-homomorphic codes} in \cite{oggierdatta} were designed with these criteria in mind.  Repair locality is particularly important for cloud storage applications, in which the disk I/O (which is proportional to the repair locality) appears to be the main judge of performance \cite{papadim}, \cite{papadimlots}.



Analyses and proposed codes for DSSs have almost exclusively concerned the logical layer, that is, the bits storing the data and their encoding and decoding.  Almost no attention has been paid to the physical layer over which the nodes communicate, and communication over this layer is usually assumed error-free.  However, the physical reality is that this channel can be noisy, for example in wireless storage systems \cite{cgong} or wireless sensor networks \cite{storagesensor}, \cite{storagesensor2}.  In \cite{cgong}, the author describes an efficient way to partially download the contents of a wireless storage network, and in \cite{storagesensor} and \cite{storagesensor2}, the authors apply network coding principles to data storage in wireless sensor networks.  Regenerating codes have been developed to correct errors in DSSs, but operate exclusively on the logical layer, as in \cite{rashmietal}.  More closely related with the present work is \cite{firststorage}, in which the authors present protocols for physical layer storage transmission encoding.

In this article, we present an analysis of the probability, $P_{\text{sub}}$, that a node subpacket is reconstructed incorrectly during node repair, for DSSs where communication between nodes takes place over potentially noisy channels.  While this is explained in more detail in Section III, we should mention now that by ``subpacket'' we mean the smallest unit into which the DSS divides the contents of a node.  In summary, the main contributions of this article are
\begin{itemize}
\item[$\bullet$] Theorem 1, which describes the asymptotic behavior of $P_{\text{sub}}$ as SNR $\rightarrow \infty$ in terms of the repair locality of the DSS and the symbol error rate of the wireless channel,
\item[$\bullet$] simulation results for AWGN and Rayleigh fading channels demonstrating how $P_{sub}$ varies with repair locality and the underlying finite field, and the accuracy of our approximation to $P_{sub}$, and
\item[$\bullet$] design criteria presented in Section VI for coding over Rayleigh fading channels in a wireless storage system, which are then used to find optimal rotations of QAM constellations which minimize $P_{\text{sub}}$ for a given repair locality, finite field size, and SNR.
\end{itemize}
In an asymptotic sense, our design criteria match those of traditional physical layer coding over Rayleigh fading channels (as in \cite{OV}), and thus it is likely that invariants such as the minimum product distance can also be useful in designing codes for the storage scenario.

\section{Basic Setup}


To establish the types of DSSs we consider, let us briefly describe a standard setup, considered for example in \cite{oggierdattabook}.  Many examples of distributed storage codes fit the following description, including the locally repairable codes of \cite{papadim} and \cite{papadimlots}, and the self-repairing homomorphic codes of \cite{oggierdatta}.  Consider a distributed storage system housing a file which is divided into $k$ pieces, distributed amongst $n$ nodes using an $(n,k)$-MDS (Maximum Distance Separable) code over $\F_q$.  This encoded data at a particular node is often subdivided into subpackets.  The MDS property guarantees that a data collector need only contact any $k$ of the $n$ nodes to collect the data.  We denote by $\omega_i$ a subpacket of the encoded data stored in the $i$th node, and we assume $\omega_i\in \F_q$ for some $q=2^m$.  

Suppose now that one of the $n$ nodes containing the subpacket $\omega$ fails, and that an incoming node has to contact $r$ helper nodes to repair the system.  The data $\omega$ can be expressed as an $\F_q$-linear combination of certain subpackets of the helper nodes.  That is, $\omega = \sum_{i = 1}^r\alpha_i\omega_i$ where the $\alpha_i\in \F_q$ are determined by the particular storage code and may depend on the particular subpacket and node which need repair.  

The $\omega_1,\ldots,\omega_r$ are to be transmitted to the newcomer via some noisy channel, for example through an AWGN or Rayleigh fading channel.  To do this, we incorporate a bijective lift function $L:\F_q\rightarrow \mathcal{C}$ which modulates the contents of the helper node into information symbols contained in a finite constellation $\mathcal{C}$ of size $q$, intended for transmission over the noisy channel.  We define $x_i:=L(\omega_i)$ for all $i$.

While many types of modulation are possible, our lift function $L$ first represents $\omega\in \F_{q}$ as a bit string of length $m = \log_2(q)$ by choosing a basis for $\F_q$ as a vector space over $\F_2$.  Then $L$ assigns to $\omega$ an element of $\mathcal{C}$, a $q$-QAM constellation, via the Gray labeling.  We will assume that all subpackets $\omega$ are uniformly distributed on $\F_q$, which guarantees that the error rates we are interested in studying are independent of the choice of basis.

The $r$ helper nodes transmit some information symbols $x_1,\ldots,x_r\in \mathcal{C}$ to the newcomer node through a noisy channel, which the newcomer decodes to get ML-estimates $\hat{x}_1,\ldots,\hat{x}_r\in\mathcal{C}$.  It then computes
\begin{equation}
\hat{\omega}_i:=L^{-1}(\hat{x}_i)\quad\text{for $i=1,\ldots,r$, and } \hat{\omega} := \sum_{i = 1}^r\alpha_i\hat{\omega}_i.
\end{equation}
The newcomer repairs the subpacket successfully if $\hat{\omega} = \omega$.  We assume that time or frequency division is employed, which allows the newcomer to receive each subpacket at a different time or frequency instant.

For a fixed channel model and modulation scheme, we define $P_s$ to be the corresponding symbol error rate.  That is, if $x$ is some information symbol transmitted over a wireless channel and $\hat{x}$ is some ML-estimate computed by the decoder, then $P_s = P(\hat{x}\neq x).$ The quantity $P_s$ of course depends on the modulation scheme and the channel.  However, we will consider only AWGN and Rayleigh fading channels in our examples, and only use $q$-QAM modulation, though many other modulations schemes are of course possible.

\section{Locally Repairable Codes}

Our motivating example of a storage code is the locally repairable codes of \cite{papadim}.  To briefly summarize their construction, a file $\mathbf{w}$ is represented as $\mathbf{w} = [\mathbf{w}^{(1)},\ldots,\mathbf{w}^{(r+1)}]$, where each $\mathbf{w}^{(i)}= (\mathbf{w}^{(i)}_1,\ldots,\mathbf{w}^{(i)}_{n})\in \F_q^n$ with $(r+1)|n$. We then define $\mathbf{s}=\sum_{i}\mathbf{w}^{(i)}$.   For our analysis it suffices to consider $r+1=n$, that is, using the vocabulary of \cite{papadim}, we only consider one ``repair group''.  When $r+1=n$, the $\mathbf{w}^{(i)}$ and $\mathbf{s}$ are then distributed among the $n$ nodes in the following way:

\begin{equation}
\begin{array}{|c|c|c|c|c|}
\hline
\text{node 1} & \text{node 2} & \cdots & \text{node $r$} & \text{node $r+1$} \\
\hline 
\mathbf{w}_1^{(1)} & \mathbf{w}_2^{(1)} & \cdots & \mathbf{w}_r^{(1)} & \mathbf{w}_{r+1}^{(1)} \\
\hline
\mathbf{w}_2^{(2)} & \mathbf{w}_3^{(2)} & \cdots & \mathbf{w}_{r+1}^{(2)} & \mathbf{w}_1^{(2)} \\
\hline
\vdots & \vdots & \vdots & \vdots & \vdots \\
\hline
\mathbf{w}_r^{(r)} & \mathbf{w}_{r+1}^{(r)} & \cdots & \mathbf{w}_{r-2}^{(r)} & \mathbf{w}_{r-1}^{(r)} \\
\hline
\mathbf{s}_{r+1} & \mathbf{s}_1 & \cdots & \mathbf{s}_{r-1} & \mathbf{s}_{r} \\
\hline
\end{array}
\end{equation}
If node $i$ is lost, a newcomer node can recover subpacket $\mathbf{w}^{(i)}_j\in \F_q$ by XORing together subpackets from all of the $r$ remaining nodes. In our notation, a subpacket $\omega_i$ can refer to any of the $\mathbf{w}^{(i)}_j$, or to the coordinate of $\mathbf{s}$ stored in the $i$th node; the analysis is independent of the particular subpacket.

The construction of \cite{papadim} uses an $(n,k)$-MDS code as an outer code, so that the $\mathbf{w}^{(i)}$ are actually themselves encoded data.  Since we are interested only in whether or not a node is repaired correctly, we will ignore the outer coding, while remarking that the MDS code allows one to potentially reconstruct the file correctly despite faulty reconstruction of individual subpackets.

\section{Subpacket Error Probability}

 Our main benchmark for judging performance will be the probability $P_{sub}= P(\hat{\omega}\neq\omega)$ that a subpacket is reconstructed incorrectly during node repair.  Let us write $\hat{\omega}_i = \omega_i + \tilde{\omega}_i$ for all $i=1,\ldots,r$, so that $\tilde{\omega}_i$ is the ``error'' introduced by the transmission of the data of helper node $i$ over the corresponding noisy channel.  Then, clearly,
\begin{equation}
P_{sub} = P(\hat{\omega} \neq \omega) = P\left( \sum_{i = 1}^r\alpha_i\tilde{\omega}_i \neq 0\right)
\end{equation}
or in other words, the subpacket is repaired correctly when the linear combination of the errors is zero.  For the locally repairable codes of \cite{papadim}, we have $\alpha_i =1$ for all $i$.  The following theorem completely describes the asymptotic behavior of $P_{\text{sub}}$ as SNR $\rightarrow \infty$, in terms of repair locality and the symbol error rate of the wireless channel.

\begin{thm}\label{bounds}
Consider a distributed storage system with repair locality $r$, where a subpacket $\omega$ is reconstructed via an $\F_q$-linear combination of subpackets from the helper nodes: $\omega = \sum_{i = 1}^r\alpha_i\omega_i$, for $\alpha_i\in \F_q$.  Then as SNR $\rightarrow\infty$ we have
\begin{equation}
\boxed{P_{\text{sub}} \sim rP_s(1-P_s)^{r-1} \sim rP_s}
\end{equation}
where $P_s$ is the symbol error rate of the channel used by the nodes.
\end{thm}

\emph{Proof:} Suppose that during node repair we have some collection $i_1,\ldots,i_j$ of $j$ out of the $r$ helper nodes which transmit their data incorrectly, that is, that $\hat{\omega}_{i_1}\neq\omega_{i_1},\ldots,\hat{\omega}_{i_j}\neq \omega_{i_j}$.  Let
\begin{equation}
c(i_1,\ldots,i_j) = \left\{\begin{array}{cl}
1 & \text{if } \alpha_{i_1}\tilde{\omega}_{i_1}+\cdots+\alpha_{i_j}\tilde{\omega}_{i_j} \neq 0 \\
0 & \text{if } \alpha_{i_1}\tilde{\omega}_{i_1}+\cdots+\alpha_{i_j}\tilde{\omega}_{i_j} = 0 
\end{array}\right.
\end{equation}
so that the quantity $c(i_1,\ldots,i_j)$ tells us whether or not the error terms cancel each other out.  Then for this fixed $j$-tuple, the quantity $P(c(i_1,\ldots,i_j)=1)$ measures the probability that the error terms will not cancel, over all possible collections of error terms $\tilde{\omega}_{i_1},\ldots\tilde{\omega}_{i_j}$.

Summing over all possible $j$-tuples of helper nodes and then over all $j$, we count all of the ways that the subpacket can be reconstructed incorrectly to arrive at
\begin{equation}\label{exact}
P_{sub} = \sum_{j=1}^r\sum_{i_1\neq\cdots\neq i_j}P_s^j(1-P_s)^{r-j}P(c(i_1,\ldots,i_j)=1)
\end{equation}
We claim that
\begin{equation}\label{boundseqn}
rP_s(1-P_s)^{r-1}\leq P_{sub} \leq \sum_{k = 1}^r\binom{r}{k}P_s^k(1-P_s)^{r-k}
\end{equation}
To arrive at the upper bound, plug the trivial inequality $P(c(i_1,\ldots,i_j)=1)\leq 1$ into the above equation, and note that there are $\binom{r}{k}$ possible $k$-tuples of helper nodes.

Now suppose that $j=1$, that is, that we are only interested in situations where the data from exactly one node, say node $i_1$, is transmitted incorrectly.  Then $\tilde{\omega}_{i_1}\neq 0$, and $\alpha_{i_1}\neq 0$ by assumption.  Therefore $c(i_1)\neq 0$, that is, there can be no cancelation of error terms when there is only one error.  Combining this fact with the trivial lower bound $0\leq P(c(i_1,\ldots,i_j)=1)$ for $j\geq 2$, one arrives at the lower bound.

We now claim that the bounds in (\ref{boundseqn}) are asymptotically tight.  That is, as SNR $\rightarrow \infty$, we claim that
\begin{equation*}
P_{sub}\sim rP_s(1-P_s)^{r-1} \text{ and } P_{sub}\sim \sum_{j = 1}^r\binom{r}{j}P_s^j(1-P_s)^{r-j}
\end{equation*}
To see this, note that by (\ref{boundseqn}) it suffices to show that
\begin{equation}\label{ratio}
\frac{\sum_{j = 1}^r\binom{r}{j}P_s^j(1-P_s)^{r-j}}{rP_s(1-P_s)^{r-1}} \rightarrow 1
\end{equation}
as SNR $\rightarrow \infty$.  To do this we treat the sum term-by-term.  For a fixed $j$, we have
\begin{equation}
\frac{\binom{r}{j}P_s^j(1-P_s)^{r-j}}{rP_s(1-P_s)^{r-1}} = \frac{1}{r}\binom{r}{j}P_s^{j-1}(1-P_s)^{1-j}\\
\end{equation}
which equals $1$ if $j=1$, and for $j\geq 2$ clearly approaches $0$ as $P_s\rightarrow 0$, i.e.\ as SNR $\rightarrow \infty$.  

The claim that $rP_s(1-P_s)^{r-1}\sim rP_s$ as SNR $\rightarrow \infty$ follows easily by expanding out $(1-P_s)^{r-1}$ and noting that the dominant term is $rP_s$. \hfill $\blacksquare$

\begin{figure}[h!]
\hspace{0em}\includegraphics[width=.45\textwidth]{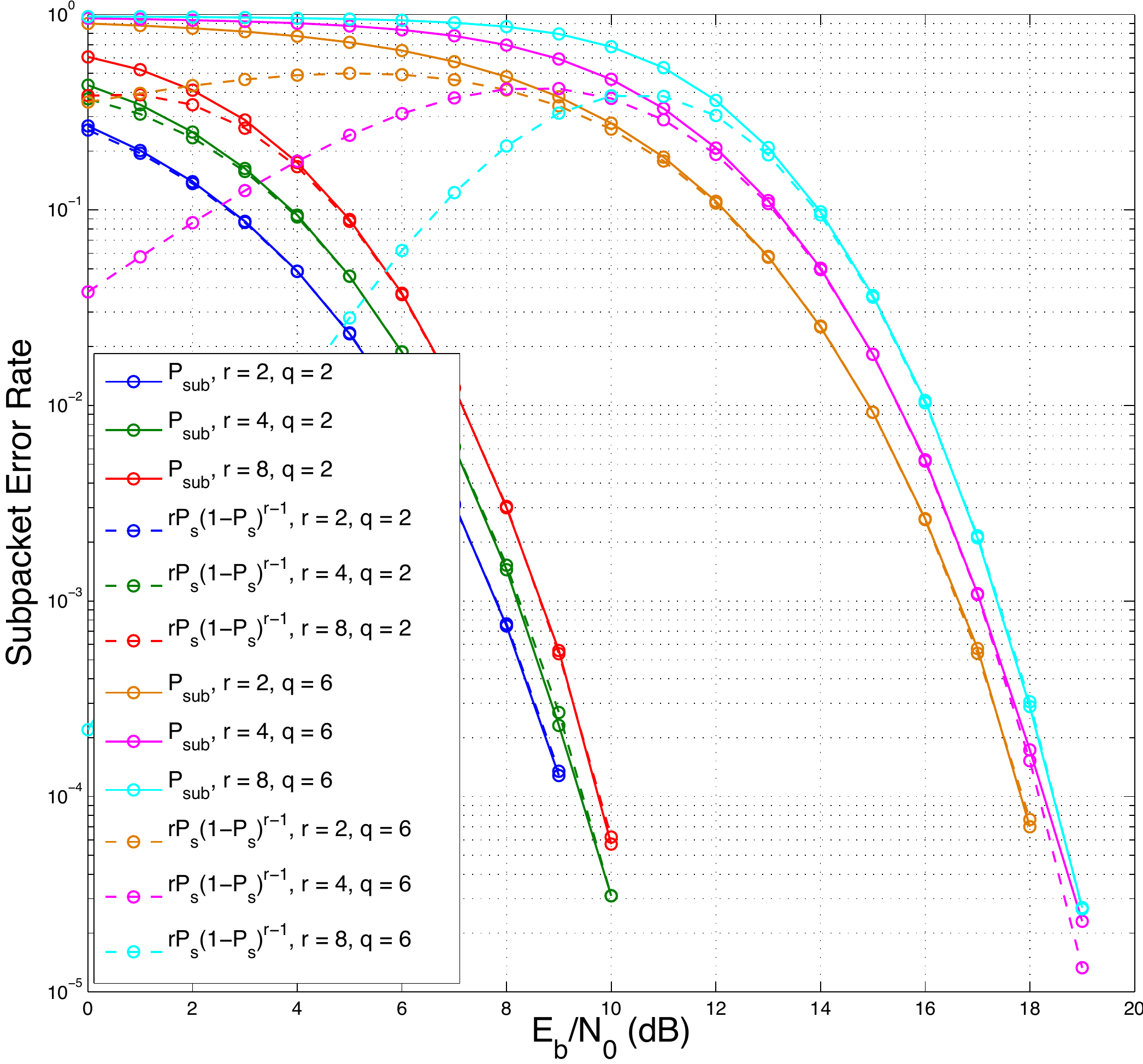}
\caption{Plots of $P_{\text{sub}}$ and $rP_s(1-P_s)^{r-1}$, for the locally repairable codes of \cite{papadim} and AWGN channel, using the finite fields $\F_{4}$ and $\F_{64}$, i.e.\ $4$-QAM and $64$-QAM.}
\end{figure}

\begin{figure}[h!]
\hspace{0em}\includegraphics[width=.45\textwidth]{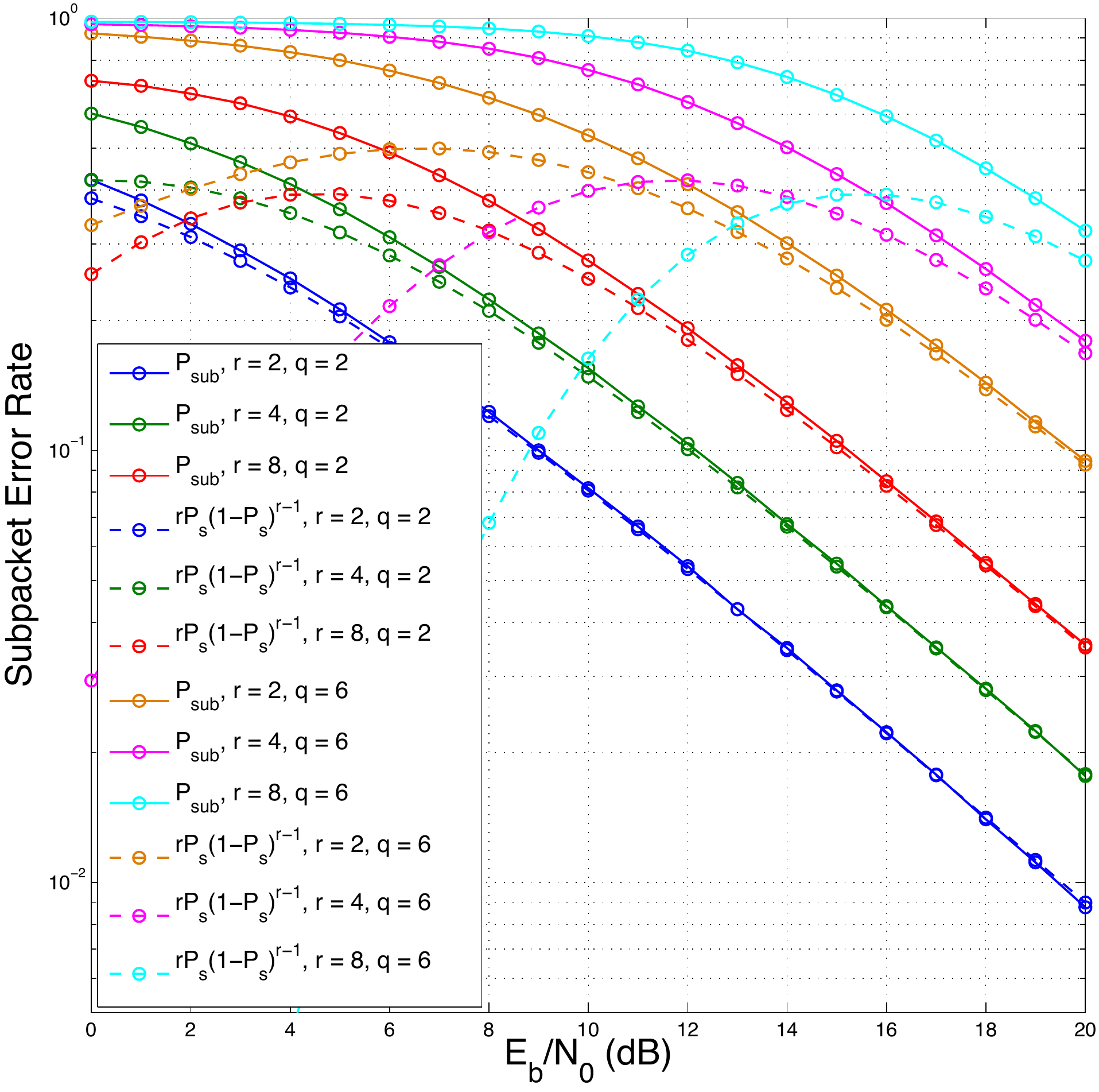}
\caption{Plots of $P_{\text{sub}}$ and $rP_s(1-P_s)^{r-1}$, for the locally repairable codes of \cite{papadim} and Rayleigh fading channel, using the finite fields $\F_{4}$ and $\F_{64}$, i.e.\ $4$-QAM and $64$-QAM.}
\end{figure}

\section{Repair over Fading Channels}

Node repair over a Rayleigh fading channel involves two equations for each helper node $i=1,\ldots,r$ (see \cite{OV}):
\begin{equation}
y_{i_j} = h_{i_j}x_{i_j} + z_{i_j},\quad \text{for $j = 1,2$}
\end{equation}
where $x_i = x_{i_1}+\sqrt{-1}x_{i_2}$ is an information symbol from a $q$-QAM constellation $\mathcal{C}$ sent by the $i$th helper node, $h_{i_j},j=1,2$ are i.i.d.\ Rayleigh random variables satisfying $\mathbf{E}(h_{i_j})^2 = 1$, $z_{i_j},j=1,2$ are i.i.d.\ zero-mean Gaussian random variables with variance $N_0/2$ representing the noise between helper node $i$ and the newcomer, and $y_{i_j},j=1,2$ are received by the incoming node from helper node $i$.  If we instead set $h_{i_j} = 1$ for all $i,j$, we are in the case of an AWGN channel.

The incoming node decodes by computing an ML-estimate $\hat{x}_i$ for each $i=1,\ldots,r$, and reconstructs $\hat{\omega}$ from the $\hat{x}_i$ as described in the previous sections.  One defines the \emph{average energy per bit} $E_b$ of the constellation $\mathcal{C}$ by
\begin{equation}
E_b = \frac{1}{\log_2(q)}\sum_{x\in\mathcal{C}}\frac{||x||^2}{q}
\end{equation}
and often plots error rates as a function of the ratio $E_b/N_0$.

To empirically test the accuracy of the approximation
\begin{equation}
P_{\text{sub}}\approx rP_s(1-P_s)^{r-1}
\end{equation}
 simulations were carried out for the AWGN and Rayleigh fading channels.   We generated $r$ bit strings $\omega_1,\ldots,\omega_r$ each of length $\log_2(q)$ uniformly at random and computed their bitwise XOR $\omega = \sum_{i=1}^r\omega_i$.  We then computed $L(\omega_i)=x_i$ using the Gray labeling, passed $x_i$ through either an AWGN or Rayleigh fading channel, decoded to get $\hat{x}_i$, and computed $\hat{\omega}$ from these.  We counted the number of subpacket errors, that is, the number of times $\hat{\omega}\neq \omega$ out of $10^6$ trials at each level of $E_b/N_0$.


In Fig.\ 1 and Fig.\ 2 we plot the results of our simulations for AWGN and Rayleigh fading channels.  One can see from the plots how the subpacket error rate increases with both the size of the underlying finite field and the repair locality.  For the AWGN channel, one can see that our approximation $P_{\text{sub}}\approx rP_s(1-P_s)^{r-1}$ is quite sharp once $P_s < 10^{-1}$.  For the Rayleigh fading channel one observes similar behavior for $4$-QAM modulation and all repair localities, and it is reasonable to extrapolate from the current data that for $64$-QAM modulation the approximation is quite accurate once $P_s < 10^{-1}$.

We omit plots describing the accuracy of the approximation $P_{\text{sub}}\approx rP_s$, since the ratio $rP_s(1-P_s)^{r-1}/(rP_s) = (1-P_s)^{r-1}$ as SNR $\rightarrow \infty$ is easily calculable, and therefore the behavior of the approximation $rP_s(1-P_s)^{r-1}\approx rP_s$ is trivial to understand.

\section{Rotation Codes for Wireless Storage Systems}

Given a distributed storage system with repair locality $r$, we can use the estimates provided by Theorem 1 to minimize $P_{\text{sub}}$.  Specifically, suppose we are using $q$-QAM modulation.  One can rotate (counter-clockwise) a $q$-QAM constellation by $\theta$ radians to improve $P_s$, and a natural question is whether or not rotations can similarly improve $P_{\text{sub}}$.

Let $\mathcal{C}$ denote a $q$-QAM constellation, and suppose that we denote by $\mathcal{C}_{\theta}$ the constellation $\mathcal{C}$ rotated by $\theta$ radians.  Let $P_s(\theta)$ and $P_{\text{sub}}(\theta)$ denote the corresponding symbol and subpacket error rates for the constellation $\mathcal{C}_{\theta}$, for a fixed finite value of SNR.  Choosing an optimal rotation demands that we solve
\begin{equation}
\arg\min_{\theta} P_{\text{sub}}(\theta), \quad \theta \in [0,\pi/2]
\end{equation}
and Theorem 1 suggests that we minimize either of:
\begin{eqnarray}
g_1(\theta) &=& rP_s(\theta)(1-P_s(\theta))^{r-1} \\
g_2(\theta) &=& rP_s(\theta)
\end{eqnarray}
Minimizing $g_2(\theta)$ is of course equivalent to minimizing $P_s(\theta)$, and there is a plethora of literature concerning optimal rotations of QAM constellations.  It is not clear for finite SNR that the same $\theta$ minimizes $g_1(\theta)$, and since $rP_s(1-P_s)^{r-1}$ is a more accurate estimate of $P_{\text{sub}}$, it is not obvious that the same rotations which are optimal for traditional QAM modulation should be optimal for the storage scenario.

Let $\mathcal{C}$ denote a $q$-QAM constellation.  It is well-known (see \cite{boutrousgood}) that the pairwise error probability for the above channel can be upper-bounded by
\[
P(x\rightarrow y) \leq \frac{1}{2}\sum_{x\neq y\in \mathcal{C}}s(x,y)
\]
where
\[
s(x,y) := \prod_{i = 1}^2\frac{1}{1+\frac{E_s}{4N_0}|x_i-y_i|^2}.
\]
In fact, if we assume that $\mathcal{C}$ has been rotated so that it is full-diversity, there is a lower bound on the pairwise error probability (see Theorem 1 of \cite{slimane}), given by
\[
P(x\rightarrow y) \geq P_L(x,y)
\]
where
\[
P_L(x,y) := \frac{1}{4}\left(\sum_{i = 1}^2\frac{1}{(1 + \delta)^i}\right)s(x,y)
\]
and
\[
\delta := \max_{i=1,2}\sqrt{\frac{\frac{E_s}{4N_0}|x_i-y_i|^2}{1+\frac{E_s}{4N_0}|x_i-y_i|^2}}.
\]
It follows that $P_s \geq \min_{x,y\in\mathcal{C}} P_L(x,y)$ and hence that
\begin{eqnarray*}
rP_s(1-P_s)^{r-1} &\leq& \frac{r}{2}\left(\sum_{x\neq y\in \mathcal{C}}s(x,y)\right) \\
&\times& \left(1-\min_{x\neq y\in\mathcal{C}}P_L(x,y)\right)^{r-1}.
\end{eqnarray*}

We can now succinctly state our two proposed design criteria as follows.  For a fixed $r$, $q$, and $N_0$, we propose selecting the rotation $\theta\in[0,\pi/2]$ which minimizes
\begin{equation}
\boxed{f_1(\theta) = \left(\sum_{x\neq y\in \mathcal{C}_{\theta}}s(x,y)\right)\left(1-\min_{x\neq y\in\mathcal{C}_{\theta}}P_L(x,y)\right)^{r-1}}
\end{equation}
or the simpler
\begin{equation}
\boxed{f_2(\theta) = \sum_{x\neq y\in \mathcal{C}_{\theta}}s(x,y)}
\end{equation}

Of course, Theorem 1 concerns asymptotic estimates of $P_{\text{sub}}$, so even though the proposed design criteria are for finite levels of SNR, we should expect rotations obtained by minimizing $f_1$ and $f_2$ to perform best for large SNR.  

\begin{figure}[h!]
\hspace{0em}\includegraphics[width=.45\textwidth]{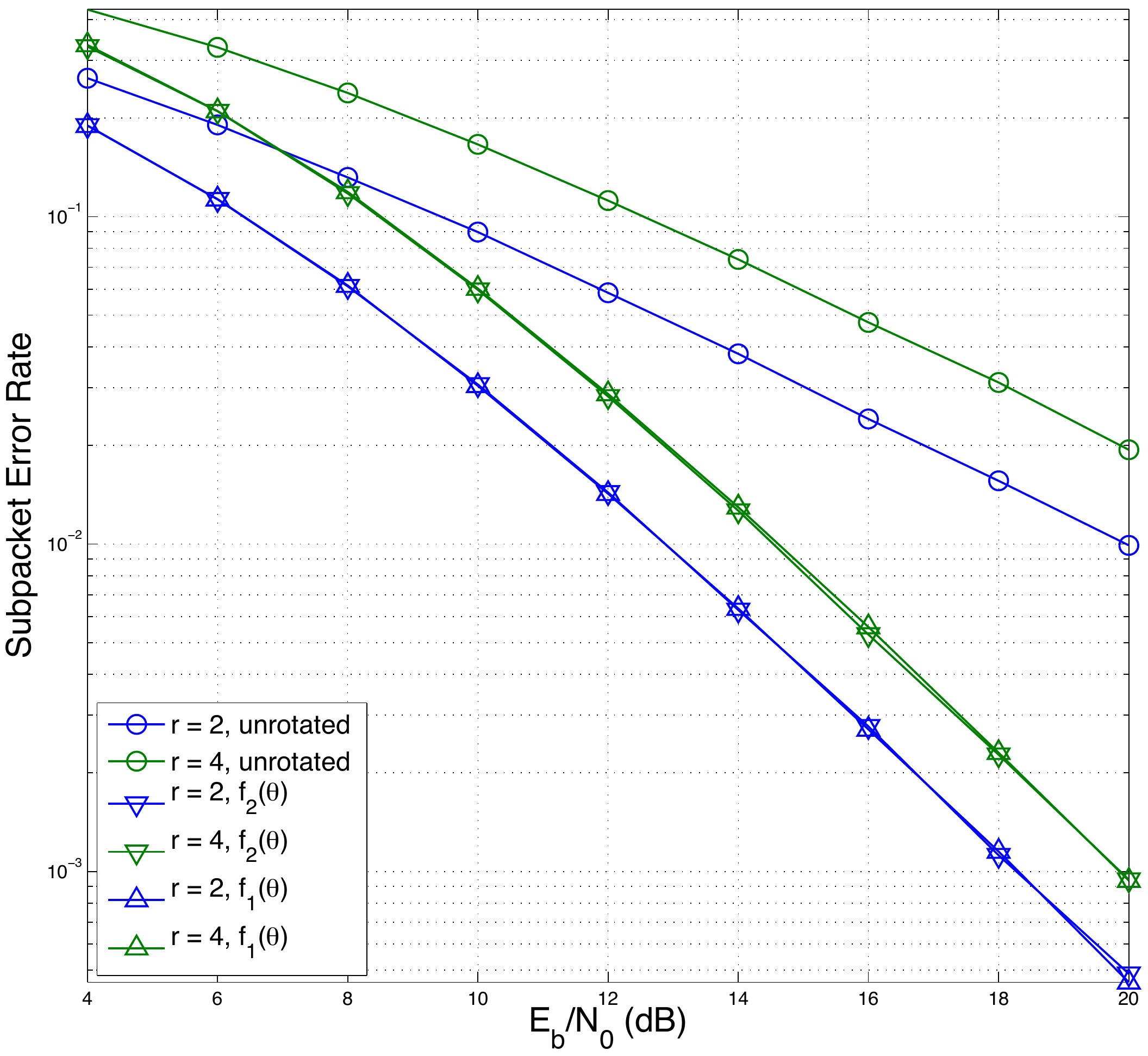}
\caption{Probability of subpacket reconstruction error for the locally repairable codes of \cite{papadim} and Rayleigh fading channel, using rotation codes for $4$-QAM constellations (i.e.\ with storage code over $\F_{16}$) optimized according to $f_1(\theta)$ and $f_2(\theta)$}
\end{figure}

\begin{figure}[h!]
\hspace{0em}\includegraphics[width=.45\textwidth]{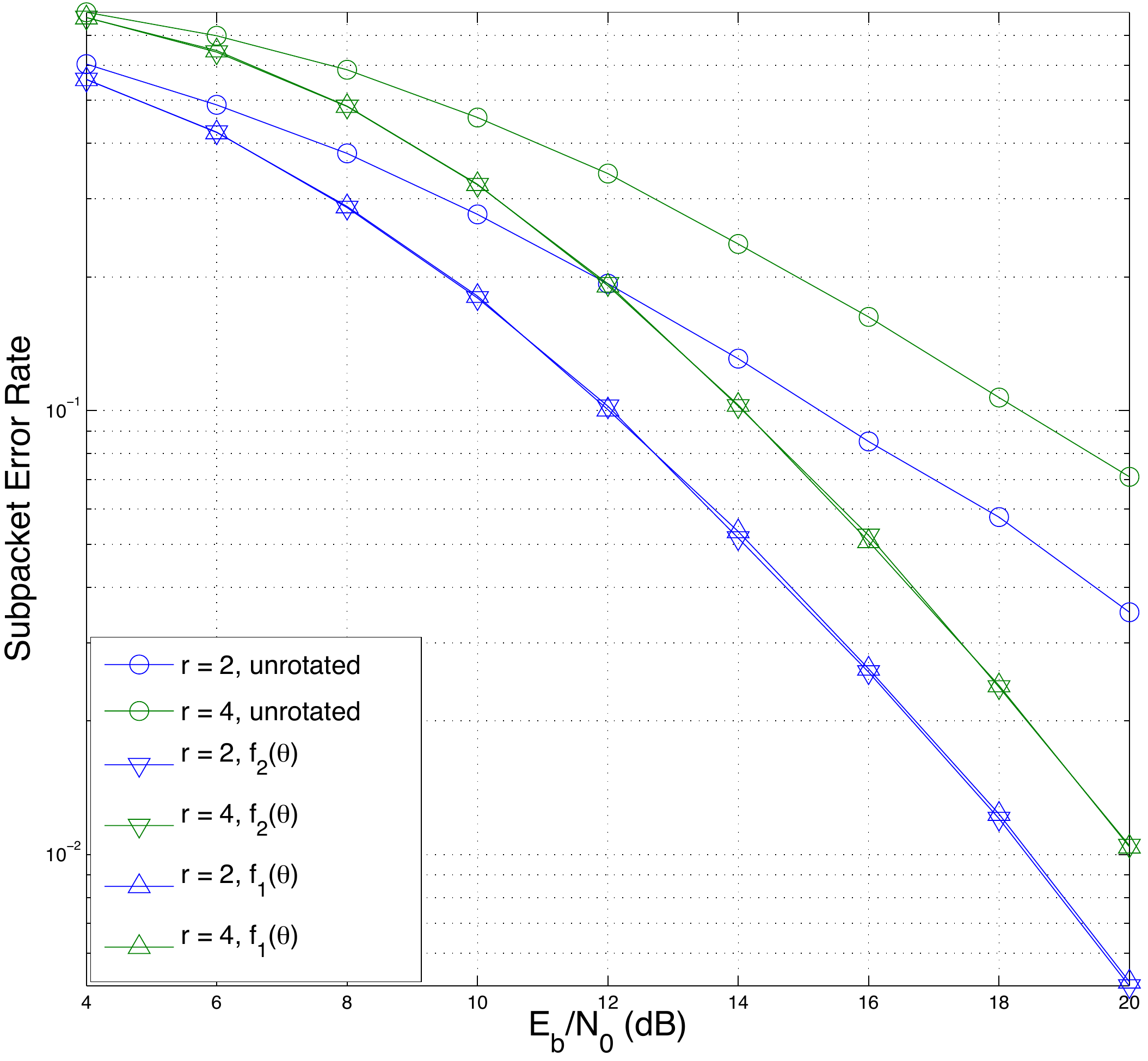}
\caption{Probability of subpacket reconstruction error for the locally repairable codes of \cite{papadim} and Rayleigh fading channel, using rotation codes for $16$-QAM constellations (i.e.\ with storage code over $\F_{16}$) optimized according to $f_1(\theta)$ and $f_2(\theta)$}
\end{figure}

To test the effectiveness of our proposed rotations, we plot in Fig.\ 3 and Fig.\ 4 the subpacket error rate as a function of $E_b/N_0$, for unrotated constellations, and for constellations rotated by the values of $\theta$ which minimize $f_1$ and $f_2$, for the given level of SNR.  Minimization of $f_1$ and $f_2$ was done by brute force, which is feasible since only one parameter determines a rotation in two-dimensional space.

It is clear from the plots that rotated constellations improve the subpacket error rate of a wireless storage system, the same way that they reduce $P_s$ for traditional $q$-QAM modulation.  Furthermore, it is also clear that the rotations produced by minimizing either objective function, namely either $f_1$ or $f_2$, have almost exactly the same performance.  In fact, nearly all of the rotations obtained by minimizing $f_1$ and $f_2$ were within $10^{-2}$ radians of the rotation for $4$-QAM which optimizes the minimum product distance.  Nevertheless, we feel it is important to establish unique design criteria for physical layer coding for wireless storage systems, because it is not obvious that the minimum product distance will continue to be good design criteria for higher dimensional lattices or wireless MIMO storage systems.

\section{Conclusions and Future Work}\label{conclusions}

In this article we described the probability that subpackets are repaired incorrectly in distributed storage systems, in which nodes communicate over AWGN and Rayleigh fading channels, in terms of the repair locality of the distributed storage system and the symbol error rate of the channel.  We used this description to establish design criteria for physical layer coding for wireless storage systems, and provided simulation results to show that our design criteria results in more accurate node repair. Further topics which deserve attention include data collection over noisy channels and generalizing our results to MIMO wireless storage systems.

\section{Acknowledgements}

The first author is supported by a grant from the Magnus Ehrnrooth Foundation and Academy of Finland grant \#268364.

\bibliographystyle{ieee}
\bibliography{myrefs_new}

\begin{thebibliography}{10}

\bibitem{papadimlots}
D.S. Papailiopoulos, J.~Luo, A.G. Dimakis, and J.~Li C.~Huang,
\newblock ``Simple regenerating codes: Network coding for cloud storage'',
\newblock in {\em IEEE INFOCOM Proceedings}, 2012.

\bibitem{oggierdatta}
F.~Oggier and A.~Datta,
\newblock ``Self-repairing homomorphic codes for distributed storage systems'',
\newblock in {\em IEEE INFOCOM Proceedings}, 2011.

\bibitem{microsoftlocality}
P.~Goplan, C.~Huang, H.~Simitci, and S.~Yekhanin,
\newblock ``On the locality of codeword symbols'',
\newblock {\em IEEE Transactions on Information Theory}, vol. 58, no. 11,
  November 2012.

\bibitem{papadim}
D.S. Papailiopoulos and A.G. Dimakis,
\newblock ``Locally repairable codes'',
\newblock in {\em IEEE International Symposium on Information Theory
  Proceedings (ISIT)}, 2012.

\bibitem{cgong}
C.~Gong,
\newblock ``On partial downloading for wireless distributed storage'',
\newblock {\em IEEE Trans. on Signal Processing}, vol. 60, pp. 3278--3288, June
  2012.

\bibitem{storagesensor}
W.~Lei, Y.~Yuwang, Z.~Wei, and L.~Wei,
\newblock ``Network coding for energy-efficient distributed storage system in
  wireless sensor networks'', April 2013,
\newblock preprint available http://arxiv.org/abs/1304.1705.

\bibitem{storagesensor2}
N.~Wang and J.~Lin,
\newblock ``Network coding for distributed data storage and continuous
  collection in wireless sensor networks'',
\newblock in {\em 4th International Conference on Wireless Communications,
  Networking and Mobile Computing (WiCOM)}, 2008.

\bibitem{rashmietal}
K.~Ramchandran P.V.~Kumar K.V.~Rashmi, N.~Shah,
\newblock ``Regenerating codes for errors and erasures in distributed
  storage'',
\newblock in {\em IEEE International Symposium on Information Theory
  Proceedings (ISIT)}, 2012.

\bibitem{firststorage}
C.~Hollanti, D.~Karpuk, A.~Barreal, and H.-F.~(Francis) Lu,
\newblock ``Space-time storage codes for wireless distributed storage
  systems'', 2014,
\newblock Global Wireless Summit, to appear.

\bibitem{OV}
F.~Oggier and E.~Viterbo,
\newblock ``Algebraic number theory and code design for rayleigh fading
  channels'',
\newblock {\em Commun. Inf. Theory}, vol. 1, no. 3, pp. 333--416, 2004.

\bibitem{oggierdattabook}
F.~Oggier and A.~Datta,
\newblock ``Coding techniques for repairability in networked distributed
  storage systems'',
\newblock {\em Foundations and Trends in Communications and Information
  Theory}, vol. 9, no. 4, pp. 383--466, 2012.

\bibitem{boutrousgood}
J.~Boutrous, E.~Viterbo, C.~Rastello, and J.-C. Belfiore,
\newblock ``Good lattice constellations for both rayleigh fading and gaussian
  channels'',
\newblock {\em IEEE Transactions on Information Theory}, vol. 12, no. 2, March
  1996.

\bibitem{slimane}
S.B. Slimane and T.~Le-Ngoc,
\newblock ``Tight bounds on the error probability of coded modulation schemes
  in rayleigh fading channels'',
\newblock {\em IEEE Trans. on Vehicular Technology}, vol. 44, pp. 121--130,
  February 1995.

\end{thebibliography}

\end{document}